\documentclass[letterpaper,twocolumn,pre, aps, superscriptaddress,floatfix,showpacs]{revtex4-1}
\usepackage{hyperref}
\usepackage{amsmath}
\usepackage{graphicx}
\usepackage{caption}
\usepackage{subcaption}
\usepackage{natbib}

\begin{document}

\title{Structure and evolution of a European Parliament via a network and correlation analysis}
\date{\today}

\author{Elena Puccio}
\email{elena.puccio@unipa.it}
\affiliation{Dipartimento di Fisica e Chimica, Universit\`a di Palermo, Viale delle Scienze, 90128 Palermo, Italy}
\author{Antti Pajala}
\affiliation{Department of Political Science, University of Turku, FI-20014 Turun yliopisto, Finland}
\author{Jyrki Piilo}
\affiliation{Turku Centre for Quantum Physics, Department of Physics and Astronomy, University of Turku, FI-20014 Turun yliopisto, Finland}
\author{Michele Tumminello}
\affiliation{Dipartimento di Scienze Economiche, Aziendali e Statistiche, Universit\`a di Palermo, Viale delle Scienze, 90128 Palermo, Italy}

\begin{abstract}
We present a study of the network of relationships among elected members of the Finnish parliament, based on a quantitative analysis of initiative co-signatures, and its evolution over 16 years. 
To understand the structure of the parliament, we constructed a statistically validated network of members, based on the similarity between the patterns of initiatives they signed. We looked for communities within the network and characterized them in terms of members' attributes, such as electoral district and party.
To gain insight on the nested structure of communities, we constructed a hierarchical tree of members from the correlation matrix. Afterwards, we studied parliament dynamics yearly, with a focus on correlations within and between parties, by also distinguishing between government and opposition. 
Finally, we investigated the role played by specific individuals, at a local level. In particular, whether they act as proponents who gather consensus, or as signers. 
Our results provide a quantitative background to current theories in political science. From a methodological point of view, our network approach has proven able to highlight both local and global features of a complex social system.
\end{abstract}



\maketitle

\section{Introduction}
Since the seminal papers by Watts and Strogatz \cite{Nature_1998}, Barabasi and Albert \cite{Science_1999}, and Newman, Watts and Strogatz \cite{PNAS_2002}, the use of networks to describe the structure and evolution of complex social systems has become a standard approach in the scientific literature. Emergent structures have been studied in social networks \cite{PRE_2003} as well as their evolution \cite{PRE_2006,PRE_2008}. 

Political systems such as the European parliaments and the US congress represent a class of social systems. The structure of similarity of the US congress has been investigated by using network theory methodologies in \cite{PhysA_2008}, where the authors built a hierarchical tree of congress members based on law initiatives they cosponsored and attempted to characterize the two biggest communities found in terms of parties (Republicans and Democrats), by using the concept of modularity. 

Although being placed in this framework, our work focuses on a typical European parliament, it takes into account more than one kind of initiatives (legislative and budget ones), and quantitatively addresses the problem of noise in a dense network, as well as the statistical significance of attributes, such as party or district, which characterize communities \cite{JStatM_2011} in the network. 

The political system under scrutiny is the Finnish parliament, which is a typical parliamentary system among its peers in Europe. In addition to the government, the parliament also has the right to initiate legislation. The thousands of private initiatives by members of the parliament have a primary signer, however, they are rather often supported by a multitude of co-signers, thus forming a network of co-sponsorships. We are interested in studying the network of these relationships among parliament members and the evolution of its structure over the years.

Our goal is to gain a quantitative understanding of the system by observing it with different techniques and by following the dynamics of its structure over time. In order to do so, we take a network approach to discover the internal ordering of the system in communities and to characterize them, and perform a correlation analysis of members, in order to study the hierarchical structure of communities \cite{Anderberg_1973, EPJB_1999, JEBO_2010} and investigate its evolution. Specifically, we study how correlations inside each party and between parties, as well as within and between government and opposition, vary over time, by using the Frobenius distance to measure how similar each year of a parliament term is to the next one. 

We look at the database of initiatives as a bipartite network in which there are two sets of nodes---parliament members on the one side, and initiatives on the other---where a parliament member is linked to an initiative if she/he signed it. According to that description, two parliament members would show a ``similar" profile if they had signed ``several" initiatives together. To provide a quantitative meaning to the expressions ``similar" and ``several", recently, a method for filtering out statistically significant links in bipartite networks has been proposed \cite{PLOS_2011_2}. This approach has already been used to investigate the structure of several systems, including stock returns \cite{QF_2015} and  investors' activity \cite{NJP_2012} in a financial market, the specializations of criminal activity \cite{PLOS_2013}, the interbank market \cite{QF_2015_2, JEDC_2015}, a mobile communication network \cite{SciRep_2014}, and a large survey on aging \cite{PLOS_2011}.

Our results indicate that the methodologies we employ are able to single out both local and global characteristics of a social system such as a parliament, which appear consistent with pre-existent theories in political science. Although our conclusions pertain to the Finnish legislature, the techniques we used are exportable to any similar systems, paying due attention to any necessary adjustments when carrying them over to a different political context.

The paper is structured as follows: in Section II we introduce datasets along with their statistics, in Section III we discuss network construction, community detection and their characterization, in Section IV we analyze correlation matrices and hierarchical trees, with a special focus on the last two terms, in Section V we present a study of the dynamics of the system, year by year, in Section VI we present results on the internal structure of the network and finally, in the Conclusions we sum up the results and point towards possible applications of our methods.

\section{Data}
The database consists of Finnish initiatives, submitted between 1999 and 2014 by members of the parliament. The information stored in the database comprises who submitted each initiative, who signed it and the year it was submitted, along with the following general attributes regarding members: their gender, party and electoral district of origin. 

Since a parliament term lasts four years, our data encompasses four different parliaments: Dataset I corresponds to the 1999-2002 parliament, Dataset II to 2003-2006, Dataset III to 2007-2010 and Dataset IV to 2011-2014. Summary statistics for each dataset are shown in TABLE \ref{tab:data}. See the Appendix for Tables of political parties with their abbreviations and number of MPs, and for the electoral districts. 

According to \cite{WPSR_2014}, the parliament members have the right to introduce a legislative initiative containing a proposal for the enactment of an Act. As the final decision in the State's annual budget lies in the hands of the parliament, parliament members can also propose a budget amendment containing a proposal for an appropriation to be included in the budget or for other budgetary decision. In addition, the parliament members can propose a petitionary motion containing a proposal for the government for drafting a law or for taking other governmental measures. Finally, a parliament member may propose to the Speaker's Council that a topical debate be held in a plenary session. All the initiatives have to be in writing and must be signed together with possible co-signatures before the introduction. Among its peers Finland belongs to the more liberal parliaments in what comes to the parliament members' private initiative introduction and there are very few restrictions regarding this activity \cite{Mattson_1995}.

The Finnish election system is proportional with open candidate lists. The country is divided into 15 electoral districts. The southern and more densely populated districts are geographically smaller than the northern ones. For example, the geographically small (capitol) Helsinki district has many more parliament members compared to the very large northern Lapland district. Finland is a parliamentary system having eight political parties represented in the parliament. During 1999-2010 two out of the three largest parties formed a majority coalition together with certain small party groups while the third large party became the main the opposition party. General elections in 2011 changed the number of large parties into four. The three largest parties formed a majority coalition leaving the fourth party into opposition together with the smaller parties. The cabinet (majority) coalition governs while the opposition has very limited power.

\begin{table}
\begin{ruledtabular}
\begin{tabular}{c | l  l  l  l } 
\multicolumn{5}{c}{\textbf {Dataset summary statistics}}  \\ \toprule
\multicolumn{1}{c}{\textit {parliament}} & \textit I & \textit {II} & \textit {III}  & \textit {IV} \\
\hline
$N$ & 2,467  & 3,163  & 3,143 & 1,808 \\
$H_i$ & 2-144 & 2-175  & 2-136 & 2-150 \\
$M$ & 179 & 186  & 183 & 199 \\
$H_m$ & 2-445 & 4-524 & 2-696 & 2-793 \\
\end{tabular}
\end{ruledtabular}
\caption {\label{tab:data} Summary statistics of data for each parliament: $N$ is the number of initiatives signed by at least 2 members, $H_i$ is the heterogeneity of initiatives, that is, the range (min-max) of signatures initiatives receive , $M$ is the number of members who signed at least 2 initiatives, $H_m$ is the heterogeneity of members, that is, the range of signatures members affix (min-max).}
\end{table}

\section{Network Description}
In this section we construct statistically validated networks (SVNs) of parliament members which point towards the presence of preferential relationships within each parliament. After having revealed the informative structure of the parliament system, we detect members' communities and find out whether these can be characterized by attributes such as party, region, and gender. The aim is to identify the key features that drive the formation of communities made of parliament members.

\subsection{Network construction}
Our system is a bipartite network, consisting of parliament members on the one side and initiatives they either proposed or signed on the other. Henceforth we treat first signers, that is, those who initially propose the initiative, simply as signers.
When dealing with bipartite networks, one can obtain a network made of nodes of the same type by projecting on the corresponding set of nodes. In our case, by projecting on the parliament members set, we obtain a network made of connections (links) between members (nodes). The projection establishes a link between each pair of members who sign the same initiative.
Unfortunately, it's reknown that such projection often produces a highly connected network, due to the heterogeneity inherent in both sets. This hides the most informative structure of the system and
therefore we apply the validation method developed in~\cite{PLOS_2011_2}. Ultimately, this resolves the validated and informative links from the random ones.
In our system, the density of links, defined as the fraction of links actually present in the nework $m$ out of the maximum number of possible links between $M$ nodes, $D=\frac{2 \ m}{M \ (M-1)}$, ranges between 0.93 and 0.99.

Before projecting on the members' set, the heterogeneity of the initiatives' set needs to be given due consideration. Indeed, initiatives display a number of signatures ranging from just a few (low degree), up to three quarters of the parliament (high degree). To account for this effect, we follow Ref.~\cite{PLOS_2011_2} and divide initiatives in bins $B_k$ according to their degree, so that each subset of the bipartite network now comprises initiatives with a specific degree range $d^i_{min}$-$d^i_{max}$ (which highly reduces heterogeneity on the initiatives' side), and only those members who actually signed them. Considering a range of degrees proves necessary for higher degrees, to keep statistical resolution high in the resulting bins. We set the binning intervals equal for all four datasets, assuming that the final results do not strongly depend on the choice made. The set of bins is defined $ B_1=\{ 2 \}, B_2=\{ 3 \}, B_3=\{ 4 \}, B_4=\{ 5 \}, B_5=\{ 6 \}, B_6=\{ 7 \}, B_7=\{ 8 \}, B_8=\{9 \}, B_9=\{ 10 \}, B_{10}=\{ 11-13 \}, B_{11}=\{ 14-20 \}, B_{12}=\{ 21-40 \}, B_{13}=\{ 41-100 \}, B_{14}=\{ >100 \}$.

If we now consider a pair of members $(i,j)$ in bin $B_k$ and each signed respectively $n_i^k$ and $n_j^k$ initiatives, out of the total number $N^k$ of initiatives within the bin, we expect to find a number $X$ of initiatives they co-signed, purely at random, that follows the hypergeometric distribution
\begin{equation}
H(X|N_k,n_i^k,n_j^k)=\frac{\binom{n_i^k}{X} \binom{N^k-n_i^k}{n_j^k-X}}{\binom{N^k}{n_j^k}}.
\end{equation}
It's straightforward then to assign a p-value to the link between $i$ and $j$ within the bin, under the hypergeometric distribution null hypothesis
\begin{equation}
p_{ij}^k=1- \sum_{X=0}^{n_{ij}^k-1}\frac{\binom{n_i^k}{X} \binom{N^k-n_i^k}{n_j^k-X}}{\binom{N^k}{n_j^k}},
\end{equation}
where $n_{ij}^k$ is the number of initiatives $i$ and $j$ co-signed, in $B_k$.  This represents the probability of randomly obtaining a value equal to or greater than what was actually observed, $n_{ij}^k$. The univariate significance level of the test, or threshold value, is usually set at either 5\% or 1\%.
In our case, we perform multiple tests, by validating the full set of links over all bins simultaneously. For this very reason, we need to cotrol for false positives, or familywise error rate, by introducing a multiple test corrected threshold. The most conservative choice, in terms of type I errors, is applying Bonferroni correction \cite{Bonf_1, Bonf_2} $\alpha_B=\alpha/m$, where $m$ is the total number of tests performed.
From here on, we choose a univariate threshold of 1\% significativity and validate each link $(i,j)$ if $p_{ij}^k < \alpha_B$, that is, if the p-value associated with the link is smaller than $\frac{0.01}{m}$, according to the Bonferroni correction for multiple comparisons. 

The last step is obtaining the weighted network. In fact, a link between the same pair of members could be validated over different bins. We account for this effect by assigning a weight to each link equal to the number of bins the link was validated in. Since we aim at building
a validated network (Bonferroni network hereafter), members who signed less than 2 initiatives, as well as initiatives with less than 2 signatures, have no relevance to our analysis and have thus been omitted. The Bonferroni network's general statistics are shown in TABLE \ref{tab:bonferroni}, along with some properties of the full network, that show the advantage accorded by this filtering method.

Basically, we're interested in studying how members cluster together on the basis of the initiatives they co-sign, thus the null hypothesis and validation allow us to distinguish information from noise.

\begin{table}
\begin{ruledtabular}
\begin{tabular}{ c | l  l  l  l}
\multicolumn{5}{c}{\bfseries Bonferroni Network statistics} \\ \toprule
\multicolumn{1}{c}{}  & \itshape I & \itshape II & \itshape III & \itshape IV \\ \hline
$M_B$ & 172 &  177 &  161 & 153\\
$m_B$ & 1633 &  1839 &  1162 & 1811\\
$f$ & 10.4\% & 10.9\% & 7.2\% & 9.9\%\\
$\mu(w)$ & 4.1 $\pm$ 2.0 & 3.6 $\pm$ 1.9 & 3.3 $\pm$ 1.9 & 2.9 $\pm$ 1.5\\ 
$\mu(w_B)$ & 1.4 $\pm$ 0.8 &  1.4 $\pm$ 0.7 &  1.6 $\pm$ 1.0  &  1.3 $\pm$ 0.7 \\
$w_m$-$w_M$  &1-8 &  1-6 &  1-9 & 1-8 \\ 
\end{tabular}
\end{ruledtabular}
\caption {\label{tab:bonferroni} 
Bonferroni network statistics for parliament terms $I-IV$. The number of members involved in the validated network $M_B$, the number of validated links $m_B$, the percentage of validated links $f$ out of the total number of original links, the mean weight $\mu(w)$ of the original network links and its s.d., the mean weight of validated links $\mu(w_B)$ along with its s.d., and links' weight range $w_m$-$w_M$ in the Bonferroni Network.}
\end{table}

\subsection{Community detection}
After building the network, our main interest lies in finding out whether it's internally organized in communities, and ultimately, which attributes, the additional information on members, characterize each community. 

Notoriously, partitioning a network is neither straightforward nor simple, there are a variety of algorithms that attempt to do so, each with its benefits and shortcomings \cite{PhysRep_2010}.
Nonetheless, the majority of techniques relies on the concept of modularity \cite{PRE_2003_2, PRE_2004}:
\begin{equation}
Q=\frac{1}{2m} \sum_{ij} \left(A_{ij}-\frac{k_i k_j}{2m} \right) \delta(C_i, C_j),
\end{equation}
where $m$ is the total number of links in the network, $A_{ij}$ is the adjacency matrix, $(k_i,k_j)$ are node i's and j's degrees, the sum is carried out over all the nodes and the delta function returns $0$ if i's and j's communities are different $C_i \neq C_j$ and $1$ otherwise.
The goal of every algorithm is to maximize the modularity over all partitions found, iteratively.

In our case, we chose a community detection software Radatools~\cite{rada00}, which employs a combination of different algorithms \cite{radatools_1,radatools_2,radatools_3,radatools_4}, allows for weighted networks and multiple repetitions of each algorithm, thus producing high modularities. We tried different combinations of algorithms, run one after the other in cascade, as detailed in the software's manual \footnote{(i) t (tabu search) r (fine-tuning by reposition) f (fine-tuning by fast algorithm) r; (ii) e (extremal otpimization) r f r; (iii) t r f r e r; (iv) e r f r t r; (v) t b (fine-tuning by bootstrapping based on tabu search) r f b r; (vi) e b r f b r.}

For each cascade, we ran several repetitions and usually had the best results by using 200 repetitions of 
(vi) e b r f b r. Furthermore, in order to check the stability of our best partition, we evaluated the similarity between the former and each of the other partitions found with the various cascades, by using the mutual information (MI) \cite{thesis_2002, entropy_2003}. As TABLE \ref{tab:detection} indicates, we found a high mean similarity, which shows a good stability of the best partition, regardless of the algorithm cascade chosen.

\begin{table}
\begin{ruledtabular}
\begin{tabular}{ c | l  l  l  l}
\multicolumn{5}{c}{\bfseries Modularity and mutual information}  \\ \toprule  
\multicolumn{1}{c}{} & \itshape I & \itshape II & \itshape III & \itshape IV \\ \hline
\itshape Q & 0.51 & 0.56 & 0.61 & 0.53\\
\itshape $\mu(MI)$ & 0.87 &  0.91 & 0.92 & 0.90\\
\end{tabular}
\end{ruledtabular}
\caption {\label{tab:detection} Modularity for the best partition $Q$ and overall mean of the mutual information between the best partition and each of the other 5 partitions, $\mu(MI)$.}
\end{table}

\subsection{Characterization of communities}
After finding a stable partition of the system, we look for each community's characterizing attributes, by validating any additional information on members contained in the database. This is accomplished in a similar fashion as what done in Section III A when validating links, following the method in \cite{JStatM_2011}. The method makes use of the hypergeometric distribution to assess the probability that a given attribute is over-expressed in the elements of a
community in respect to all the elements of the investigated set. Again, each attribute is validated, community-wise, if the associated p-value falls below the Bonferroni threshold at 1\% of significativity.

The attributes we consider are both specific, such as a member's party, district and gender, and more generic, as whether the district was in a rural or metropolitan area (Helsinki and Uusimaa), whether the party was in the government or in the opposition and its political position (right, centre or left).

Overall, we found a satisfactory characterization of communities by party, district or both, as shown in Fig.~\ref{fig:comm}, and some characterization by area, coalition and political position (details in the appendix, TABLES~\ref{tab:government} - \ref{tab:term_IV}). Surprisingly, gender doesn't appear characterizing, indicating there is no influence of a member's gender over whom he collaborates with. 
\begin{figure}
    \includegraphics[width=0.5\textwidth]{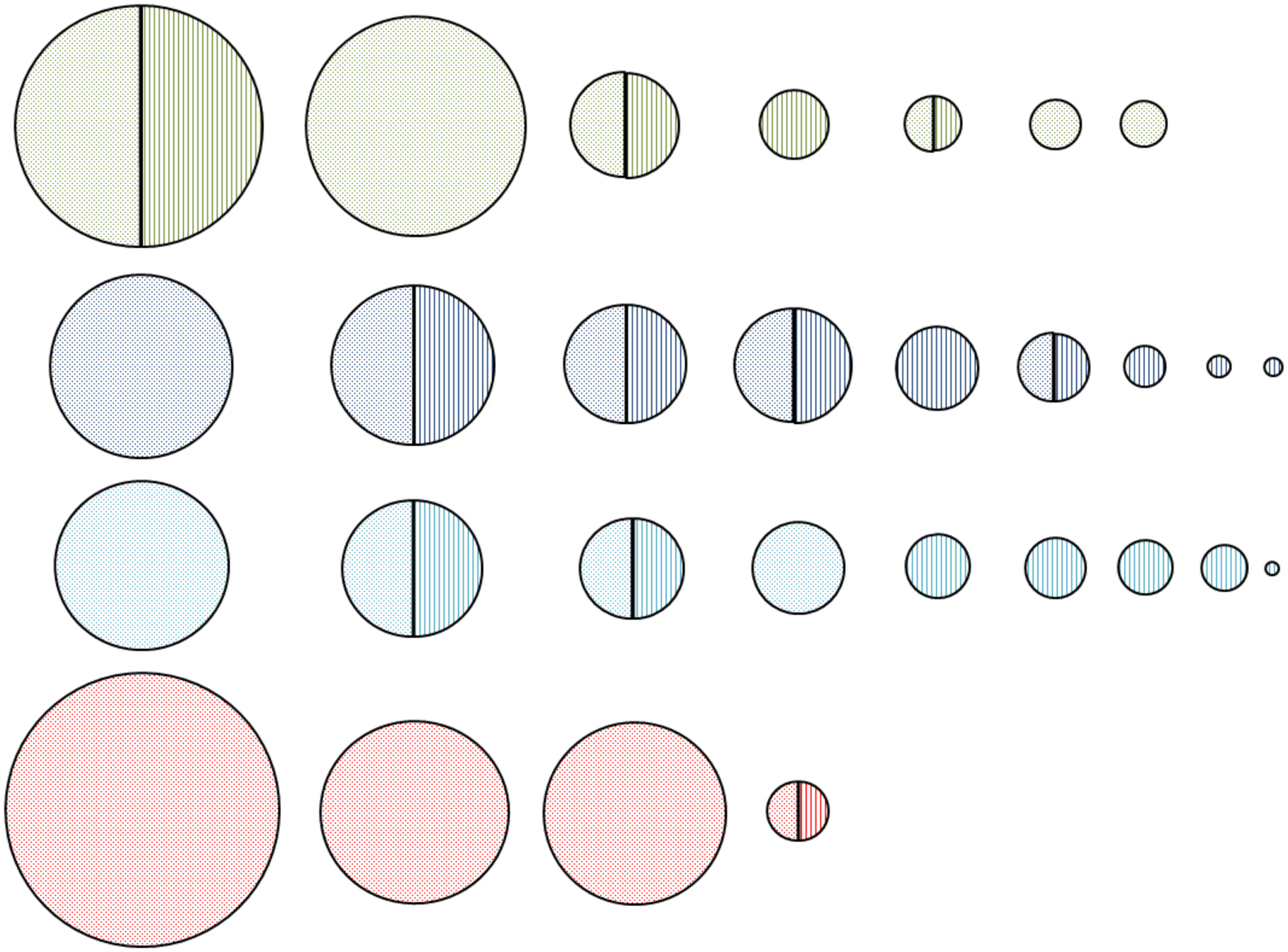}
    \caption{
    Qualitative illustration of communities. For full information, see TABLES IV-VIII in the Appendix. Top row with 7 communities is the I parliament term (green), II term has 9 communities (blue), III has 9 (cyan) and IV has 4 (red). A dotted pattern indicates characterization by party, a striped pattern characterization by district, the presence of both patterns indicates characterization by both. The radius of each circle is proportional to the number of members within the community. For example, parliament II in the second row has a large number of modest size communities where all except one has district characterization. In contrast, for parliament IV (fourth line), the number of communities is limited but their size is large, with only one community characterized by district whilst party drives community formation.}
     \label{fig:comm}
   \end{figure}

Still, the most interesting result is that, although party appears more explanatory for the first dataset, showing up in all communities, both party and district concur to characterize communities in the second and third datasets, with district growing in importance. Finally, party reverts to being the most characterizing attribute for the last dataset. 

\begin{figure}
    \includegraphics[width=0.4\textwidth]{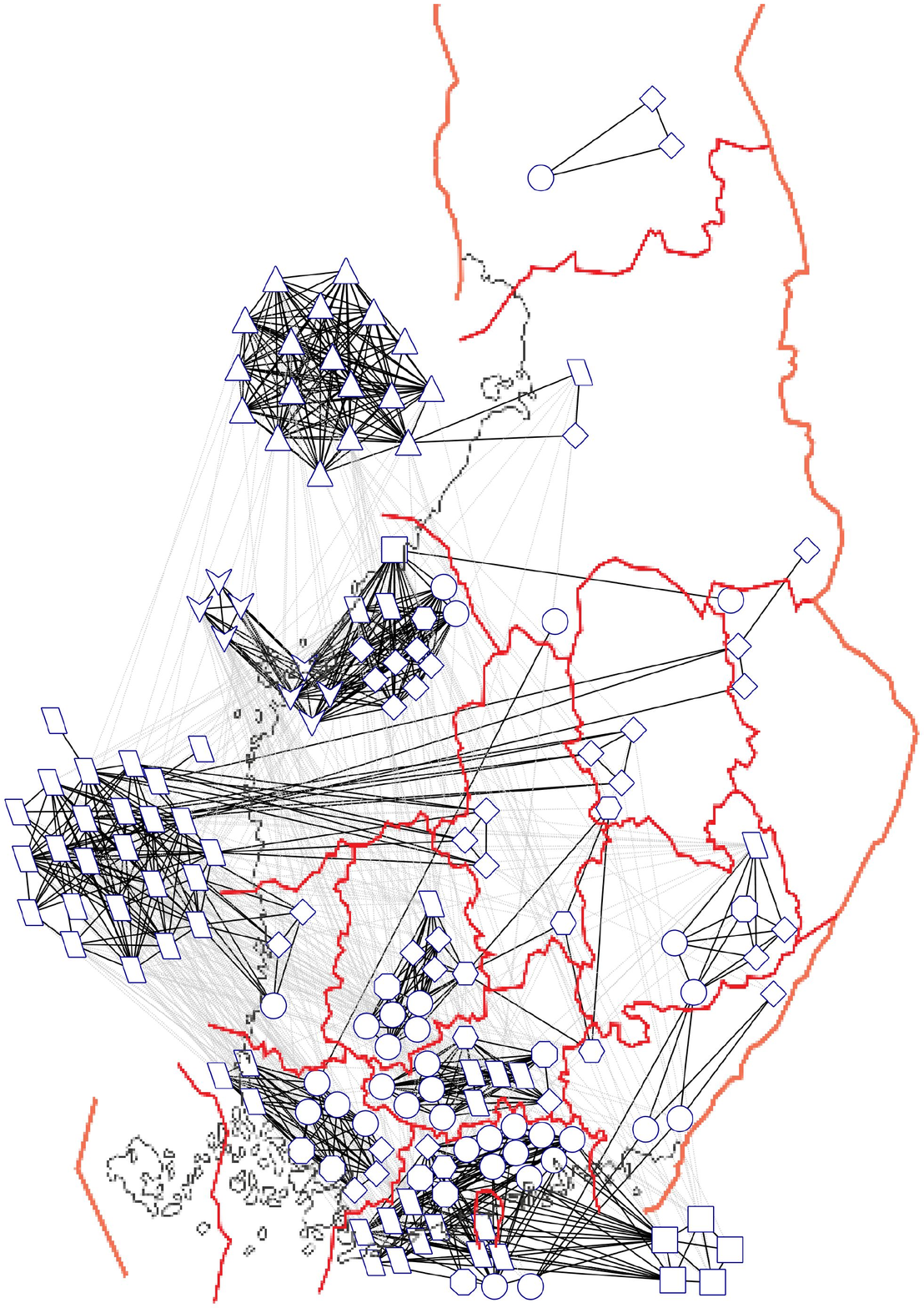}
    \caption{Bonferroni Network for the III parliament. The map in red lines shows electoral districts in Finland, node shapes indicate each member's party: hexagon=KD, diamond=KESK, ellipse=KOK, rectangle=PS, V=RKP, parallelogram=SDP, triangle=VAS, octagon=VIHR. Nodes are located either on the map, according to their district of origin or outside the map, when party characterization dominates (the network reflects what shown in Fig.~\ref{fig:comm}, 3rd row). To emphasize the network partition, links within each community are displayed in black, links between communities are in light gray. All communities are well characterized by district, with the exception of parties VAS and SDP, on the left-hand side. }
    \label{fig:map}
\end{figure}

Indeed, it appears that collaboration moved from being among members of the same party to being among members from the same district, as time passed by, with the last parliament going against the trend. The Bonferroni Network for the third parliament is shown in Fig. ~\ref{fig:map}, this is the best example of districts playing a major role in how members cooperate.
Broadly speaking, the power of our method lies in that it allows us to quantitatively characterize communities in the network and is further able to reveal changes occurring in the structure of the collaboration web among members, from term to term.

\section{Correlation analysis}
In this section, we investigate the structure of the correlation matrix with the aim of building a hierarchical tree of members. The advantage that the correlation matrix offers over the network approach, is the hierarchical ordering of members. This allows to look for substructure within each cluster. The drawback is that we no longer use a quantitative tool to discriminate the random part of correlations present in the system.

The first step to obtain a correlation matrix between members, is to define a correlation coefficient, that hinges on the similarity between patterns of signatures. Specifically, we'd like to evaluate the similarity between a pair of members $(i, j)$ by using a correlation coefficient $ s_{i,j} \in [-1,1]$ \cite{PRE_2015} that takes into account the number of joint signatures $(n_{ij})$, the number of individual signatures $(n_i, n_j)$ and the total number of initiatives $(M)$. Basically, when considering binary vectors, that is, vectors whose entries can be either 1 or 0, this turns out to be just a straightforward derivation of the Pearson correlation coefficient
\begin{equation}
 s_{i,j}=\frac{n_{ij}-n_i n_j/M}{\sqrt{n_i(1-n_i/M)n_j(1-n_j/M)}}.
\label{eq:similarity}
\end{equation}

\subsection{Hierarchical trees}
Hierarchical trees are built by using the average linkage method, that is, by successively merging pairs of clusters $A$ and $B$ according to a mean distance between elements $x \in A$ and $y \in B$ given by
\begin{equation}
\mu(d_{x,y})=\frac{1}{|A| \cdot |B|} \sum_{x \in A} \sum_{y \in B} d_{x,y},
\end{equation}
where $|A|$ and $|B|$ indicate the number of elements belonging to cluster $A$ and $B$, respectively. The quantity $d_{x,y}$ is a measure of distance between members $x$ and $y$ \cite{EPJB_1999}, which is a function of the correlation coefficient given in Eq. \eqref{eq:similarity}:
\begin{equation}
d_{x,y}=\sqrt{2 \ (1- s_{x,y})}.
\end{equation}

By looking at all trees, we again find that both parties and districts play a role in how members cluster together, within each parliament term. Such a strong characterization by party, district and coalition gives an insight on what actually determines members affiliation to a specific cluster.

\begin{figure}
    \includegraphics[width=0.5\textwidth]{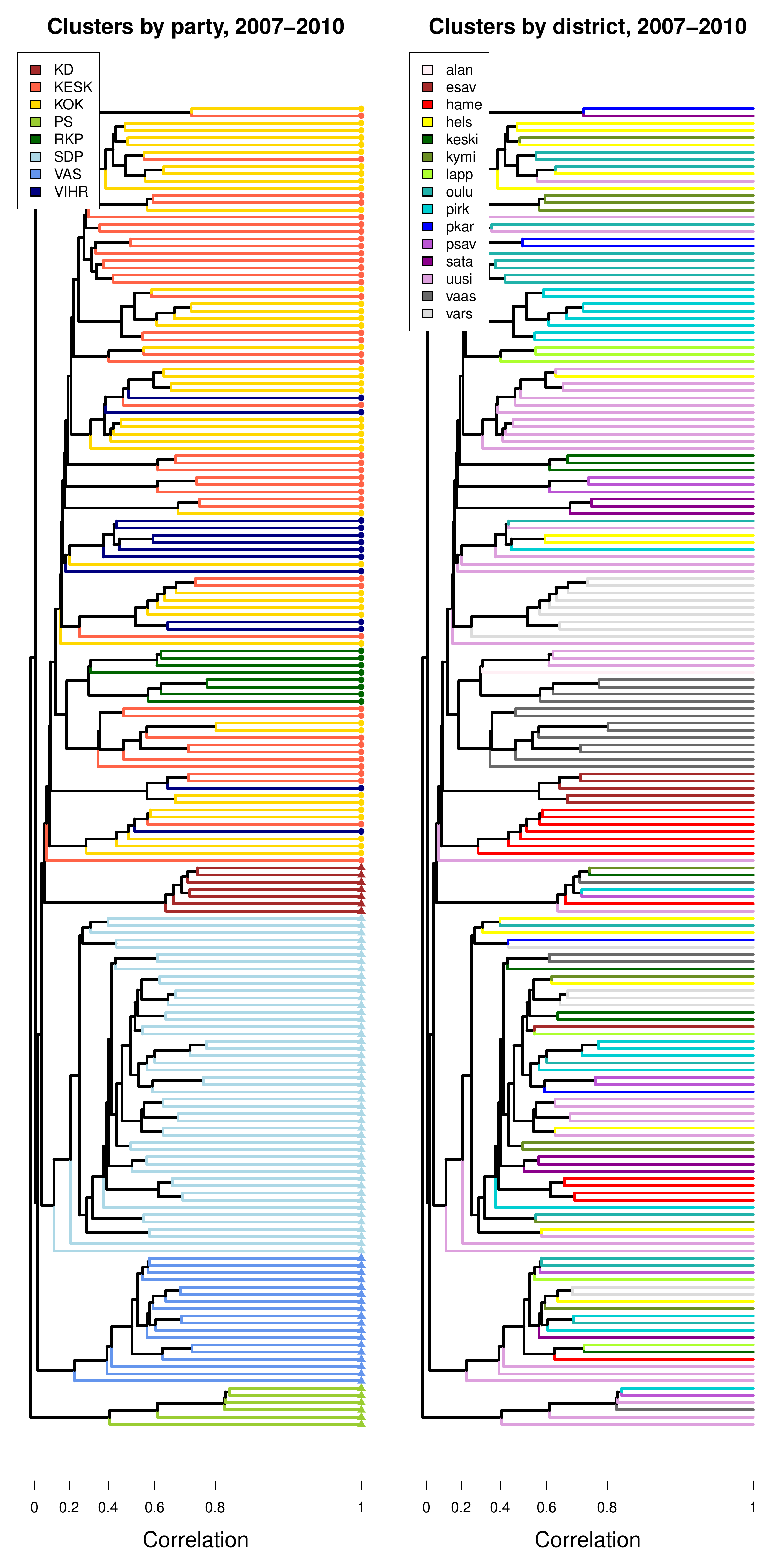}
    \caption{Clustering of members during the III parliament term. On the left, leaves and branches are colored according to parties, circle-shaped leaves indicate nodes in the government (top half of the tree), while triangles indicate nodes in the opposition (bottom half). Coloring on the right-side tree is according to district. The x-axis shows correlation values, going from 0 (no correlation) to 1 (maximum correlation), the spacing is not linear due to the measure of distance chosen. Party dominates in the opposition, while district sub-clustering occurs in the government.}
    \label{fig:dendro3}
\end{figure}

\begin{figure}
    \includegraphics[width=0.5\textwidth]{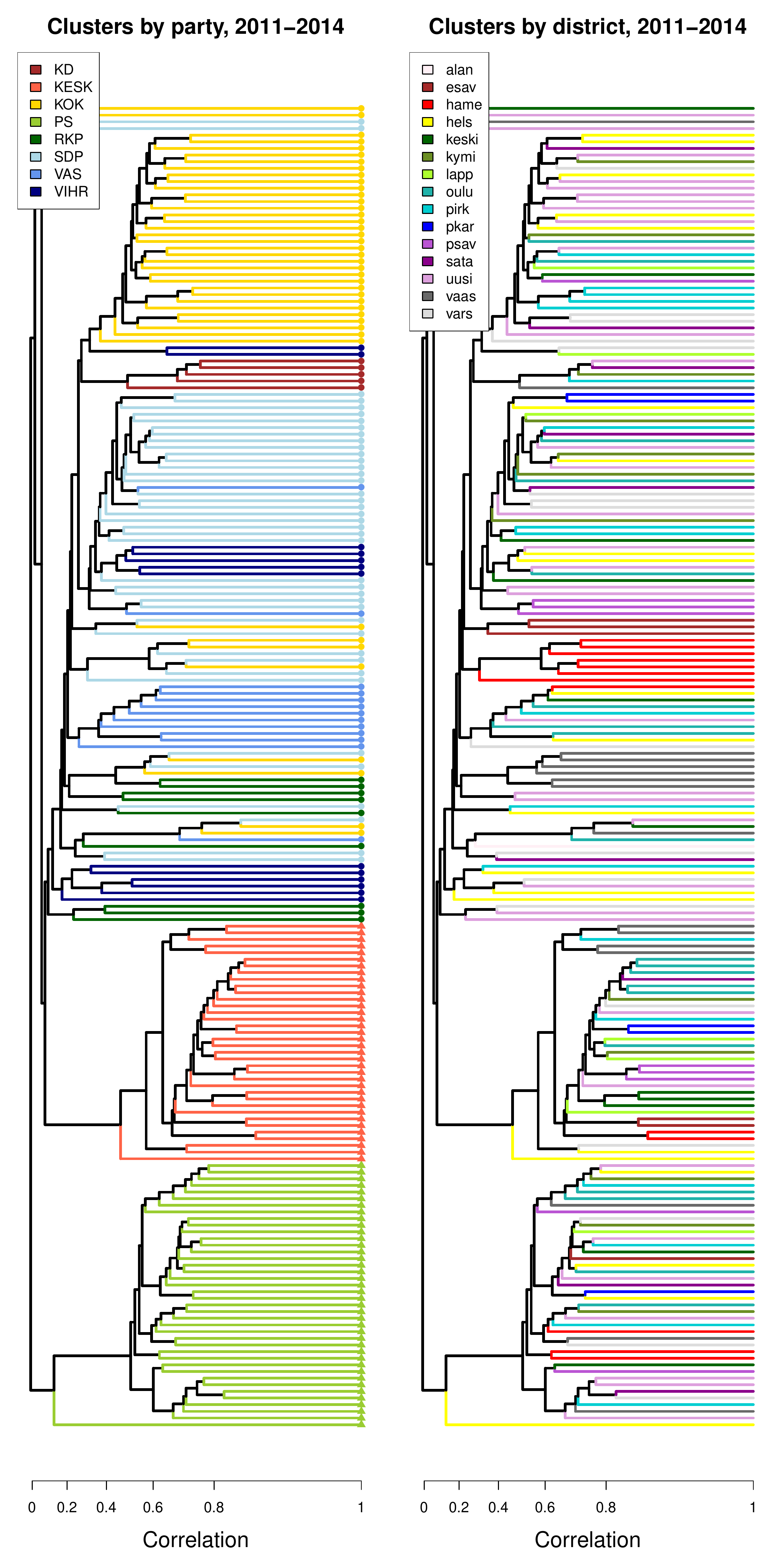}
    \caption{Clustering of members during the IV parliament term. On the left, leaves and branches are colored according to parties, circle-shaped leaves indicate nodes in the government (top half of the tree), while triangles indicate nodes in the opposition (bottom half). Coloring on the right-side tree is according to district. The x-axis shows correlation values, going from 0 (no correlation) to 1 (maximum correlation), the spacing is not linear due to the measure of distance chosen. Here party dominates strongly both in the opposition and in the government, with sporadic subclustering by district. Noticeably, the two opposition parties do not cluster together.}
    \label{fig:dendro4}
\end{figure}

The main result lies in the hierarchical structure revealed by this method: the opposition tends to cluster strongly by party, while the government shows a collaboration among parties and a further subclustering by district. We chose to display figures just for the last two terms, where the change in the system's structure is more striking, see Figs.~\ref{fig:dendro3} and~\ref{fig:dendro4}.

Political position doesn't appear to be crucial, in general left-wing parties are somewhat closer together. On the other hand, coalition plays a major role, with government and opposition parties neatly separated. Specifically, opposition parties show closed ranks, while the government cluster doesn't show much of a party structure. Finally, some districts seem to work together, in particular, the metropolitan ones of Uusimaa and Helsinki. This behaviour is common to the first three parliaments, see Fig.~\ref{fig:dendro3} for the hierarchical tree of the third term. An exception to this overall behaviour is represented by the last term, where we found the government coalition clustering mainly by party against the trend of the previous years, as shown in Fig.~\ref{fig:dendro4}.

\subsection{Structure of the correlation matrix}
\begin{figure}
    \includegraphics[width=0.5\textwidth]{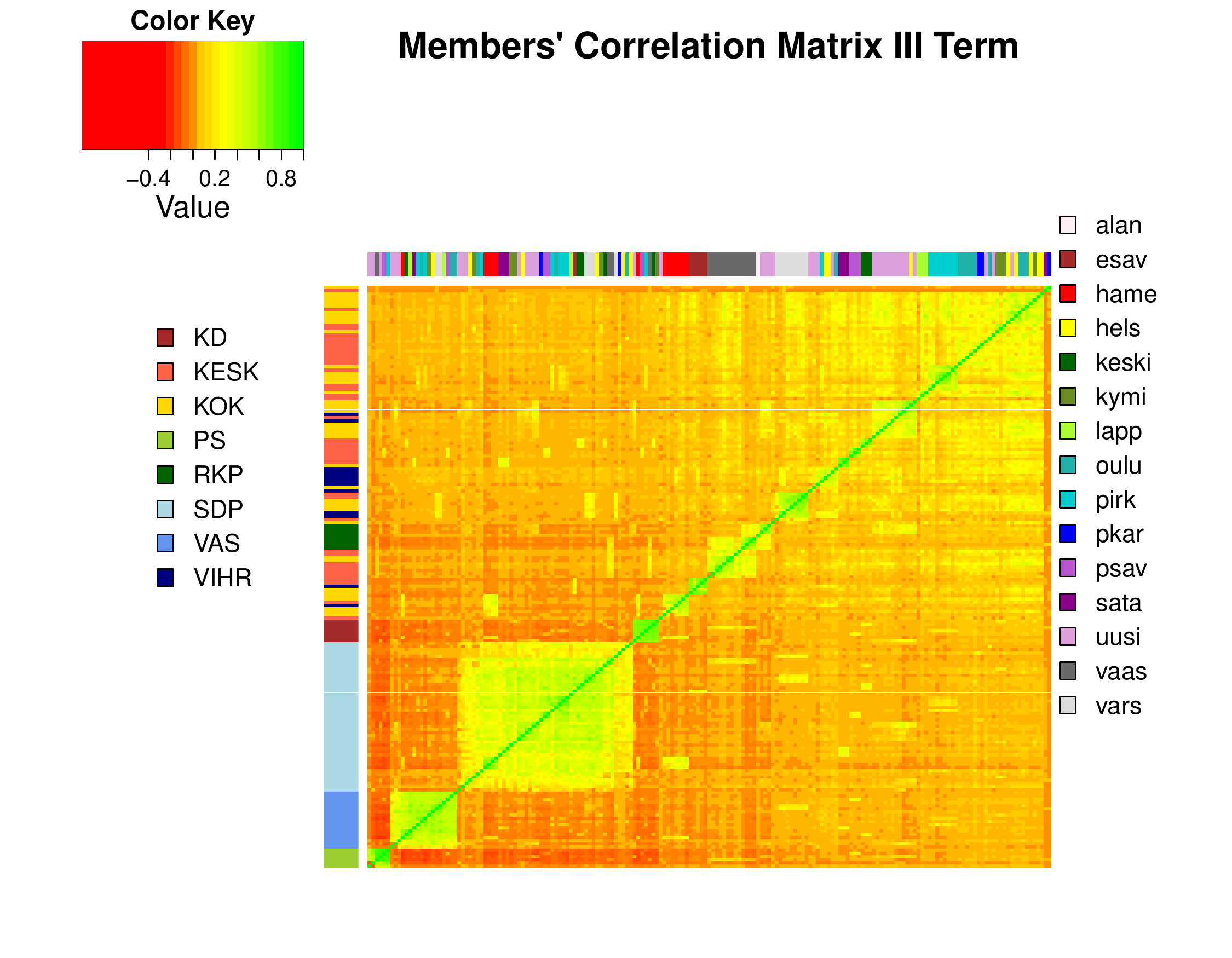}
    \includegraphics[width=0.5\textwidth]{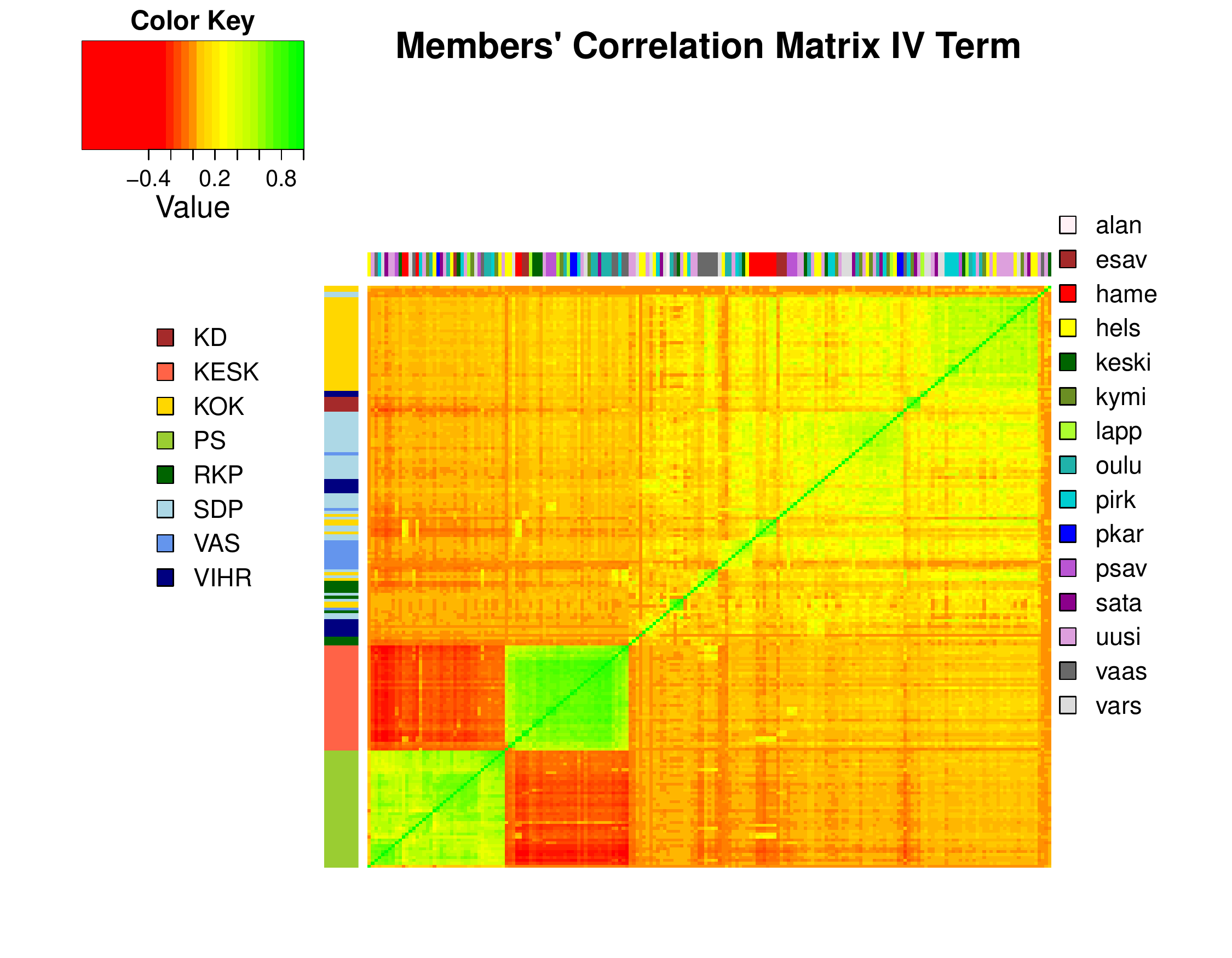}
\caption{\label{fig:correlation} Correlation matrices of members for III (top panel) and IV (bottom panel) parliament terms.  
The color key displays negative correlations in red, near-zero values in orange, mild correlations in yellow and high correlations in green. Colored bars on the side indicate either members' party (left bar) or district (top bar), as displayed by side legends. In each matrix, members are ordered according to each term's correponding hierarchical tree.
In the III term, district sublclustering (colored bar on top) can be clearly discerned within the government (top-right square), from left to right: hame, esav, vaas, vars, sata, psav, keski, the metropolitan districts of uusi and hels, lapp and pirk. Party clustering is present in the opposition (bottom-left 4 green clusters corresponding to parties: ps, vas, sdp and kd).
In the IV term, we find party subclustering within the government (top-right square), with right-wing parties at the top and left-wing parties at the bottom. The 2 opposition parties (ps and kesk) are on the lower-left corner and anti-correlate with each other. Here we notice how district subclustering almost disappears.}
\end{figure}

In this section, we focus on the correlation matrices for the last two parliaments (see Fig.~\ref{fig:correlation}), where the subclusters are easily seen, along with the change of structure from one term to the next. Members are arranged according to the hierarchical trees in the previous subsection.
Apparently, during the III term there's a higher overall correlation and small district-clusters in the government, along with 4 bigger party-clusters in the opposition. On the other hand, during the last term, a collaborative government is clearly seen as a yellow huge cluster with a left and right subdivision and some party-subclusters. For the opposition, there are now only 2 very compact parties, who slightly collaborate with the government and anti-correlate with each other.

By comparing the general structure of the correlation matrices, we draw the conclusion that party counts more than district during the last parliament and opposition parties make a statement of behaving in opposite ways, which is noticeably different than previous terms. Our approch is thus able to detect the change that occurs in the system's structure when the parliament changes, and to quantitatively assess the different webs of collaboration/anti-collaboration that arise within it.

\section{Some Dynamical Features Within and Over the parliamentary terms}
We focus now on more detailed look on the dynamical features, in particular by looking at the annual evolution of the correlation matrix.
Correlation matrices are henceforth computed year by year, and arranged in blocks representing parties, in order to have a closer look at how interactions evolve within each party (diagonal blocks) and between them (off-diagonal blocks).

\subsection{Mean correlations}
We are interested in how correlations within each party and among parties evolve year by year. The purpose here is obtaining a closer look at the dynamics of interactions between members depending on the party they belong to. We consider the mean correlation calculated over members belonging to a specific party (intra-party), and the mean correlation calculated between members of that party and each of the others (inter-party).

\begin{figure}
    \includegraphics[width=0.5\textwidth]{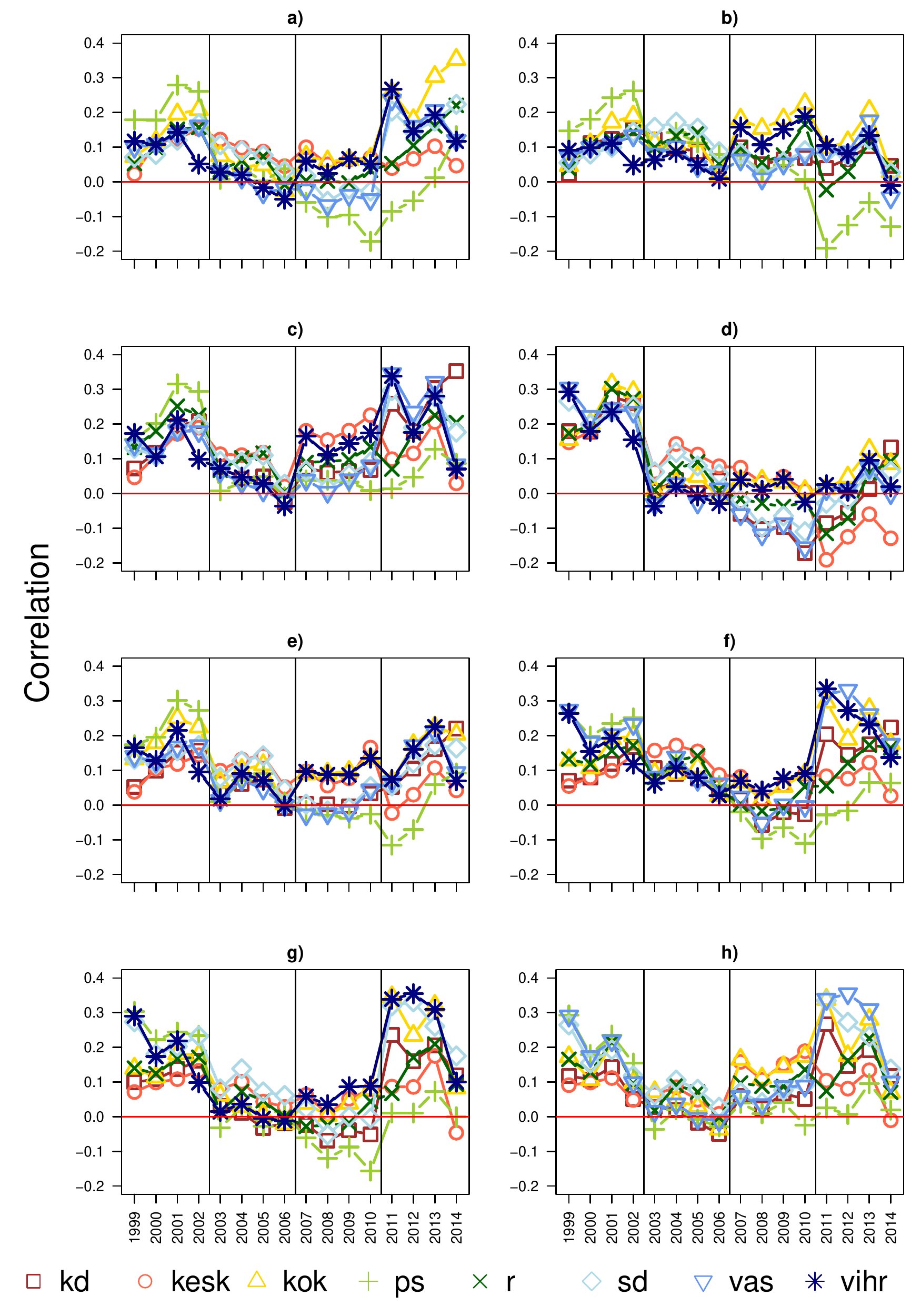}
    \caption{Mean correlation between parties, over time. Panel a) Inter-party mean correlation between KD and all the others; b) KESK-others; c) KOK-others; d) PS-others; e) RKP-others; f) SDP-others; g) VAS-others; h) VIHR-others. Red horizontal lines mark the 0-value. Vertical lines separate parliament terms. The general trend shows higher inter-party mean correlation during I term, decreasing, packed together, going below zero, mean correlation values during II and III term and increasing, broader values during last term (with the sole exception of the 2 opposition parties).}
    \label{fig:mean_cor}
\end{figure}
\begin{figure}
    \includegraphics[width=0.5\textwidth]{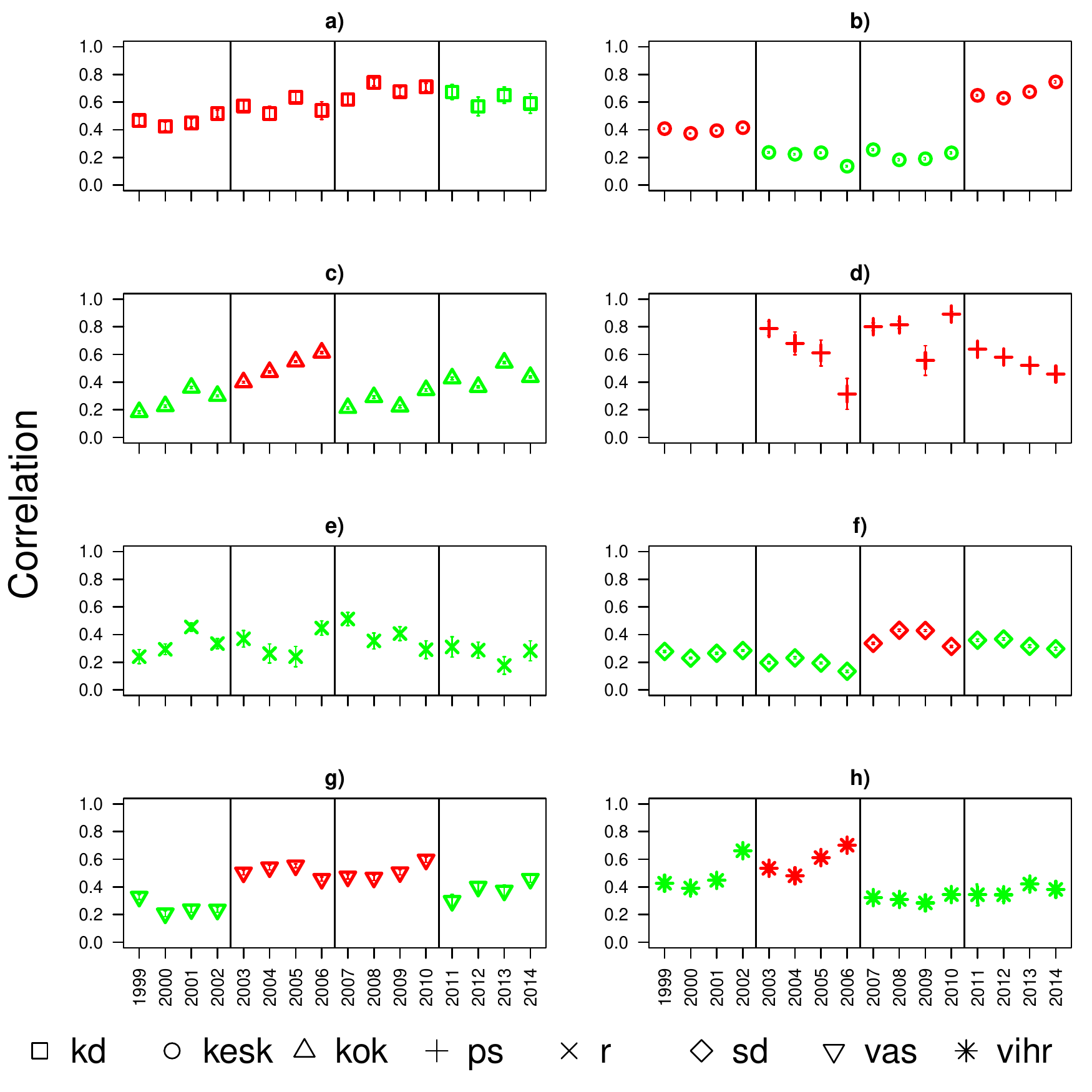}
    \caption{Mean correlation within each party, over time. Panel a) Intra-party mean correlation for party KD; b) KESK; c) KOK; d) PS; e) RKP; f) SDP; g) VAS; h) VIHR. Vertical lines help distinguishing each parliament term. In general, the trend indicates that when parties are at the government (shown in green) they have lower mean correlation than when at the opposition (shown in red). In any case, intra-party mean correlation values are on average higher than inter-party ones, as expected. Error bars are 6 standard deviations of the mean.}
    \label{fig:mean_cor_intra}
\end{figure}

From Fig.~\ref{fig:mean_cor} and~\ref{fig:mean_cor_intra}, we can see how each party presents its own pattern of collaboration with every other party, which is not constant over the years, not even within a single parliament term. Nonetheless, there are some general trends, for example it's apparent how the first and even more so the last term are unusual in the sense of having spread out of correlations, while second and third terms (in the middle) display a narrower, more similar pattern. Moreover, mean inter-party correlation tend to be positive during the I term, still positive but closer to zero during the II, they turn slightly negative for opposition parties during the III (kd, ps, sdp, vas) and plunge even more for opposition parties during last term (kesk, ps). Error bars were calculated by bootstrapping data 1000 times and choosing 5-95\% confidence intervals, and are all of the order of $10^{-2}$.

The overall mean correlation between a party and all the others, all parliaments considered, is strictly positive, ranging from 0.06 to 0.12. So, on average, ps is the least collaborative, while kok, sdp and vihr seem the most willing to collaborate, as expected from parties often at the government. Therefore, we can see the behavioural change of parties, especially in correspondance to a change of parliament.

\begin{figure}
    \includegraphics[width=0.5\textwidth]{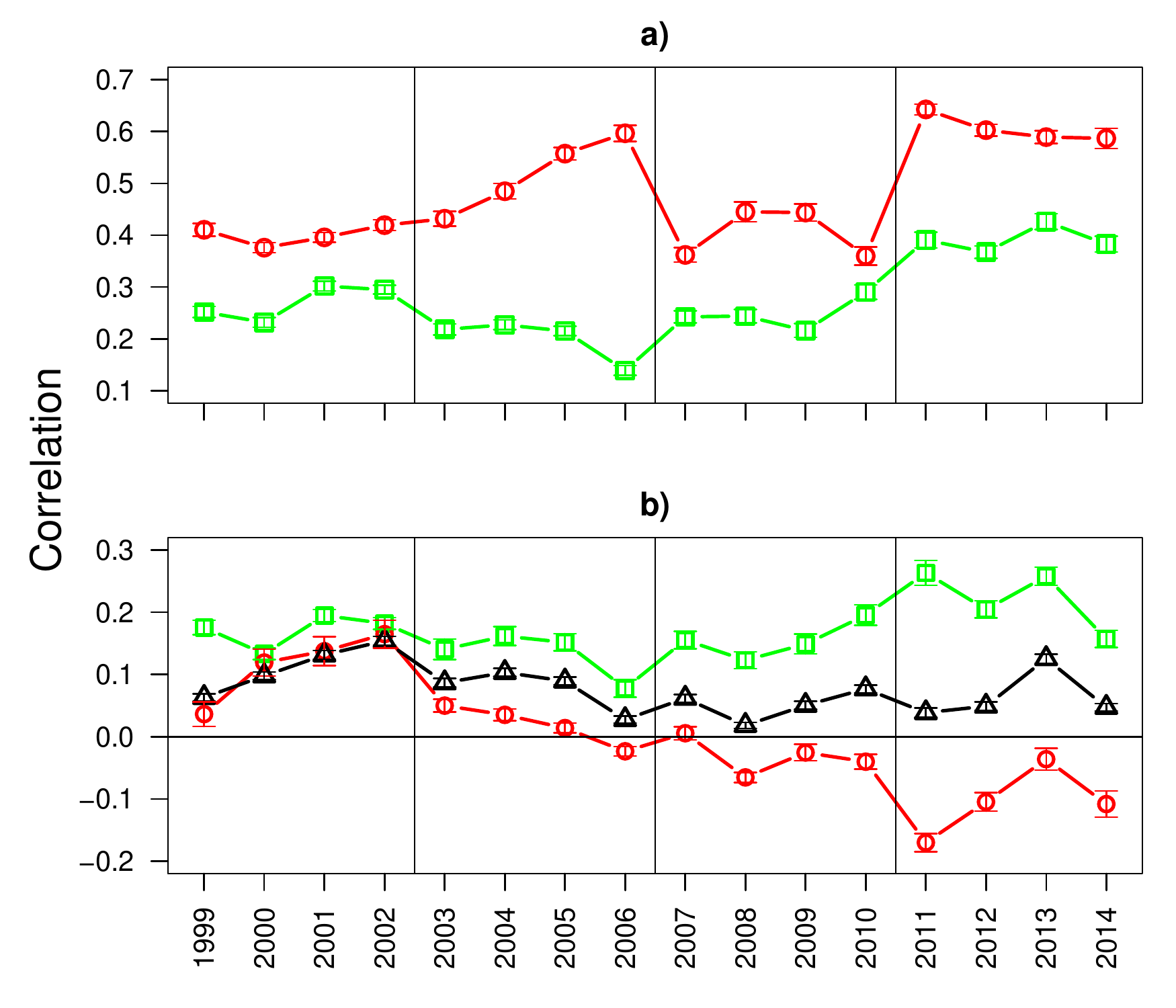}
    \caption{Panel a): Mean intra-party correlation for government parties (green squares), and for opposition parties (red circles), over time. Opposition parties display a higher mean correlation throughout the parliament terms.
Panel b): mean inter-party correlation between all government parties (green squares), between all opposition parties (red circles) and between government and opposition parties (black triangles), over time. Here correlations between government parties become noticeably higher over time than those between government and opposition, which still best those between opposition parties. The 3 curves tend to spread out over time, and again collaboration between government parties is high, whilst that between opposition parties grows scarcer and scarcer.
Error bars are 6 standard deviations of the mean. Vertical lines separate terms, and the 0-value is clearly indicated.}
    \label{fig:correlations}
\end{figure}

To get a closer look at how correlations vary within government and opposition coalitions, we repeated the whole analysis by grouping parties according to their coalition. Our main finding is that mean correlation tends to be higher within parties in the opposition than parties in the government, while collaboration is higher between parties in the government than between those in the opposition, as expected (see Fig.~\ref{fig:correlations}). Collaboration between government and opposition is low, but still better than collaboration within the opposition, which becomes negative by the end of the second parliament, stays slightly negative during the III and takes a dive during the fourth. When looking at mean inter-party correlation we also notice how the trend is for the 3 curves to spread out over time.

\begin{figure}
  \includegraphics[width=\linewidth]{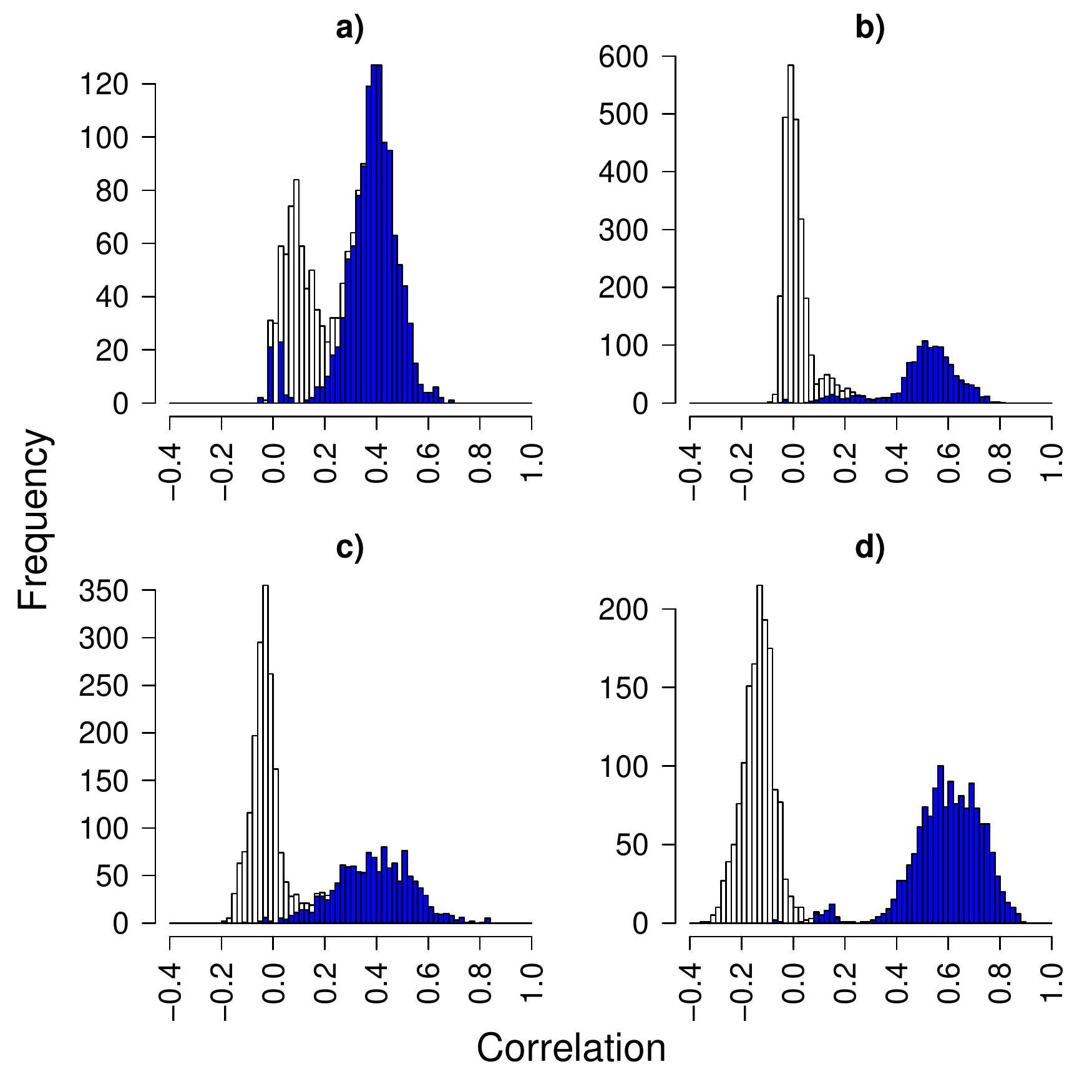}
\caption{Distribution of correlations between members of opposition parties. Panel a) I parliament; b) II parliament; c) III parliament; d) IV parliament. The shaded area (in blue) represents correlations within each party, the rest are correlations between opposition parties. All distributions are two-peaked, showing how collaboration is strong within each party (right peak), but weak between different opposition parties (left peak), this feature grows stronger from term to term and is enhanced during the last one.}
\label{fig:hist}
\end{figure}

In order to stress how high values of the mean correlation within the opposition do not reflect a collaboration between parties, we also looked at the distribution of correlations within the opposition. What we found is that these high values within the opposition are only due to high intra-party correlations, but do not point towards a strong collaboration between parties, as shown in Fig.~\ref{fig:hist}. In fact, the distribution of correlations is not homogeneous, but displays 2 peaks. By coloring the area due to intra-party correlations, we can explicitly see how indeed these fall on the high side (right peak), while correlations between opposition parties (left peak) center on near-zero values for the first 3 terms and turn negative during the last term.

On the other hand, if we look at the distribution of correlations within the government, we find that it is rather homogeneous, which is an indication of how government parties present more of a united stance.

\subsection{Annual distance within each parliament}
In this section we study the Frobenius distance between the correlation matrices of members corresponding to the 4 different years composing each parliament. We aim to understand whether there are similar years within a parliament term and unusual ones, in terms of how members cooperate as expressed by the correlation matrix relative to each year.

The Frobenius distance between two matrices $\Sigma_1$ and $\Sigma_2$ is defined as:
\begin{equation}
F(\Sigma_1, \Sigma_2)=\sqrt{Tr[(\Sigma_1-\Sigma_2)(\Sigma_1-\Sigma_2)^T]}
\end{equation}

In order to compare two matrices using this measure, their dimension must be the same. For this reason, we restricted the analysis to members who signed at least 1 initiative per year (and thus are active every year).

\begin{figure}
    \includegraphics[width=0.5\textwidth]{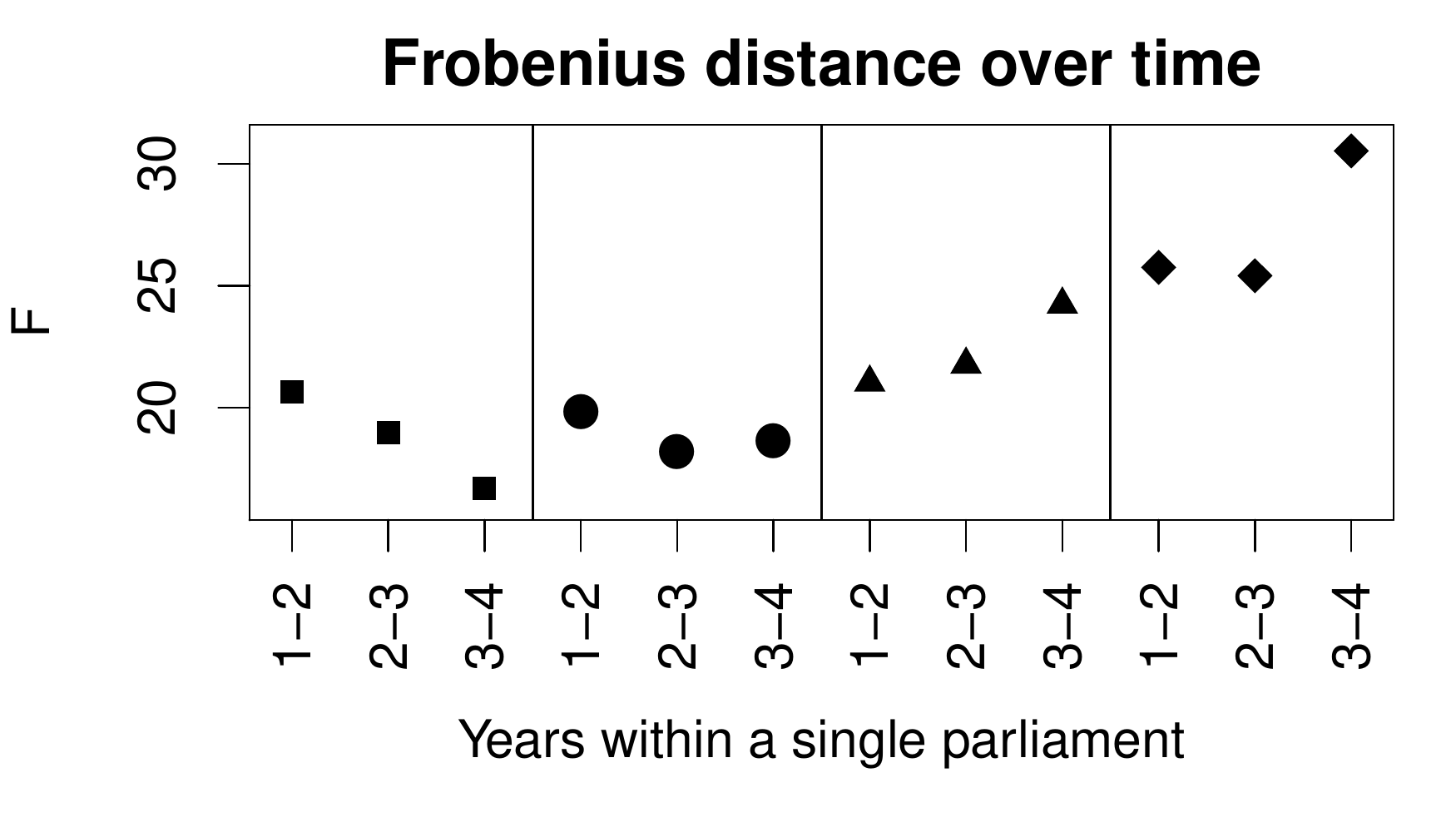}
    \caption{Frobenius distance between correlation matrices corresponding to a year and the one immediately after, within each parliament. Different symbols indicate different parliaments, separeted by vertical black lines: squares for the I term, circles for the II, triangles for the III and diamonds for the last one.}
    \label{fig:frob}
\end{figure}

The results are shown in Fig.~\ref{fig:frob} for all terms, although it should be borne in mind that it's not possible to compare parliaments, because the corresponding correlation matrices involve different members, are in general of different dimensions, and the Frobenius distance depends on the dimension of the matrices involved.

From Fig.~\ref{fig:frob}, we can see an initial stabilising trend, followed by a diverging one. Indeed, for the first term, the most anomalous year is the first, when the parliament has just formed, while the last 2 years are the most similar. The second term is in a sort of stable equilibrium, while for the last two terms, the trend is diverging, with the 3rd and 4th years most dissimilar and the first two being more or less alike. 


\section{Internal Structure: Reciprocity and Disparity}
We turn our attention now on individual parliament members. Specifically, we are looking for special nodes, as for example members of relevance for their own party (local leaders), or members who gain favour from many others (influential people), as well as members who mostly sign other's initiatives (followers).

\subsection{Reciprocity}
A feature of our social system we're interested in is reciprocity, or in other words the tendency of a member to reciprocate a signature he received.

As a measure of reciprocity between two people, we suggest the following:

\begin{equation*}
r_{i,j}= \frac {n_{i \rightarrow j}-n_{j \rightarrow i}}{n_{i \rightarrow j}+n_{j \rightarrow i}} , \ \ \ i<j
\end{equation*}

where $n_{i \rightarrow j}$ stands for the number of signatures by $i$ on initiatives proposed by $j$ and $n_{j \rightarrow i}$ is vice-versa. The two extreme categories are the one with $r=0$, implying full reciprocity, and the one with $r= \pm 1$, implying unreciprocation of signatures (either from $i$ to $j$ or from $j$ to $i$). Interestingly, most members fall in these two categories, meaning that either they fully reciprocate all signatures they receive ($r=0$), or they just gain signatures from others without ever once returning them ($r= \pm 1$).

In our data we can single out a few members in each term who receive an outstanding amount of signatures (from several hundreds, to a few thousands) from all the others, that they do not reciprocate. The Table~\ref{tab:reciprocity} in the Appendix shows the top three scores of reciprocity. 

\subsection{Disparity}
An interesting feature of the internal structure of our network is, more in general, the presence of net givers and net receivers of signatures. We introduced four centrality measures, that aim to weigh the tendency of a member to sign others' initiatives against his tendency to receive signatures. Basically, we're introducing ``disparity'' measures of out-degree/in-degree imbalance in our bipartite network.

The measures we consider take into account the number of initiatives proposed and signed by member $i$, the number of signatures member $i$ received, the number of different members who signed $i$'s initiatives and the number of different members whose initiatives $i$ signed. Unfortunately, given the several dimensions of heterogeneity involved, there is no unique way to calculate out-degree/in-degree imbalance for our system.

The first measure we consider is the number of initiatives $n_p$ proposed by member $i$ minus the total number of initiatives $i$ signed, $n_s$, normalized to the sum:
\begin{equation*}
d_1= \frac {n_{p}-n_{s}}{n_{p}+n_{s}}.
\end{equation*}
If $d_1$ is positive, it indicates that member $i$ tends to propose more initiatives than he signs: he's more focused on submitting proposals than on signing any.

The second measure is the number of signatures $i$ received on average per initiative proposed $n_r/n_p$, minus the number of initiatives $i$ signed, normalized to the sum:
\begin{equation*}
d_2= \frac {n_r/n_p-n_{s}}{n_r/n_p+n_{s}}.
\end{equation*}
If $d_2$ is positive, it means that member $i$ receives more signatures per initiative than he affixes. Whoever scores high here, proposed just a few very successful initiatives, without signing many initiatives proposed by others.

The third measure is the absolute number $n_r$ of signatures $i$ received minus the number of initiatives $i$ signed, normalized to the sum:
\begin{equation*}
d_3= \frac {n_r-n_{s}}{n_r+n_{s}}.
\end{equation*}
A positive value of $d_3$ implies a member is a net receiver overall, regardless of the number of initiatives he proposed.

The last measure is a bit different from the previous three in that it focuses on members instead of signatures. It counts the number of different members $m_r$ who signed $i$'s initiatives minus the number of different members $m_s$ whose initiatives $i$ signed, normalized to the sum:
\begin{equation*}
d_4= \frac {m_r-m_s}{m_r+m_s}.
\end{equation*}
This measure, $d_4$, is a direct proxy of whether a member $i$ has a wide network of collaboration: a high (low) score here indicates a member is a global receiver (signer), linking members from different clusters.

Tables~\ref{tab:disparity_top} and~\ref{tab:disparity_worst} in the Appendix show the top and worst scores, respectively, for all these measures. 
Looking at who comes out on top, in terms of the number of unreciprocated signatures received and the scores on the four disparity measures, and considering also the lowest scores, we get the following picture, consistent over all terms: there are few highly active members who rank top, the leaders, and a bulk of members who rank low and are mainly signers (followers).

Opposition parties are the most active, directing initiatives through one or two selected members who are not party leaders, but often are veterans or chairmen, while the rest of the party, the signers, simply sign their initiatives. This is easily seen by looking at the maximum number of unreciprocated signatures received, at measures $d_1$ (high number of proposals) and $d_3$ (high absolute number of signatures received), where we find the same couple of members from opposition parties, showing a huge level of activity, along with many members of government parties who display a moderate activity.

Government parties dominate measures $d_2$ and $d_4$, where the first number represents the mean number of signatures received per proposal, while the second number measures the number of signatures by different members. This is probably due to the way government parties collaborate among them by both proposing and signing each other's initiatives, and in so doing they display a larger area of influence.

Although the highest scores appear dominated by the government, with just a few very active members in the opposition, acting for their own parties, the lowest scores are completely ruled by the opposition. The reason for this is, as we stated, that the whole party focuses support for its spokesmen, trying its best to push their plans through, since they can't count on any substantial collaboration outside their own party.

The scenario that emerges from this analysis is one of a government with a stronger collaboration web between parties, producing a certain number of moderately active people who split the job of proposing and signing initiatives among themselves, widening their network of collaborations beyond just their party. On the other hand, opposition parties present a strong front, channeling proposals through one or two members and raising consensus within their own party.

\section{Conclusions}

The statistically validated network method has been used to retrieve the informative structure of a social system, the Finnish parliament. After reducing the density of links in our network to the fundamental core, according to the Bonferroni correction for multiple comparisons, we validate communities and find that not only party drives the network of collaboration between members of the parliament, but also their district of origin. Network characterization changes over time, as the parliament changes, due to both a partial change in its components (roughly 40\%-50\% of members remain the same from one parliament to the next), and to an evolution of the collaboration network itself. In particular, a tipping point in this evolution is represented by the outburst of the PS party in 2011, which determined an apparent increase of the importance of party in characterizing communities of members and a subsequent decrease of the role played by the district of election. 

The statistically validated network analysis does not allow one to gain deep insight about either the behavior of members who show peculiar patterns of signatures, because they turn out to be excluded from the network (about 10\% of parliament members), or the sub-structure of revealed communities. To gain a closer insight into the nested structure of communities, how they emerge and how collaboration profiles change from parliament to parliament, we perform a correlation analysis and construct hierarchical trees of parliament members based on pairwise similarity. We show that the structure of similarity of parliament members change from a term to the next, due to the varying composition of the parliament itself, i.e., for instance, the proportion of members between two parties change from a term to the next. However, a major change of the structure of the system occurs between the third an fourth parliament. Indeed the increased number of parliament members from the PS party in the last term, not only changed dramatically the composition of the parliament at the relative weight of the different parties, but it also polarized the behavior of parliament members, who became less prone to cooperation outside of their party. Such an inter-party collaboration, though, is more pronounced among parties that support the government, and such a stylized fact is persistent throughout the sixteen years analyzed in the paper. The result is that government and opposition show an intrinsically different behavior in that within the former, collaboration happens across party lines, with distinct sub-clusters characterized by district of election, whilst, within the opposition, parties close ranks and tend not to collaborate with each other. As a result, different parties in the opposition can even become negatively correlated, as is the case during the last parliament. Such a behavior of opposition parties is so extreme that they tend to collaborate more with the parties that support the government than with each other.

The correlation analysis also shows how the parliament inner workings change, according to the Frobenius distance between correlation matrices of one year and the next within a single term. Specifically during the first term we observe a negative trend of the Frobenius distance over the years, indicating that the correlation structure of the parliament tends to stabilize over time. Such a stabilizing trend almost disappears in the second parliament, where the similarity between the correlation matrices is stable over time. In the third parliament, we observe an inversion of the trend, and correlation matrices at the end of the term are more different than those at the beginning. Finally, such a de-stabilizing trend becomes very pronounced in the last parliament. 

To study the system also from the perspective of single parliament members, and to observe how their influence changes over time, we develop ad hoc measures, and look at  parliament members as leaders, influential groups and followers. The result is again a different behavioral pattern between opposition, which channels all initiatives through one or two leaders and a bulk of followers, and government, which develops a wider collaboration web, displaying a variety of influential members who split the load between them.

To sum up, we developed and used methodologies that prove themselves effective when studying multiparty democracies or parliamentary systems, as are those within the EU, especially so when there are many factors concurring to determine a network of collaboration. 

\acknowledgments
J.P. acknowledges financial support from Magnus Ehrnrooth Foundation. E.P. thanks the Turku groups for hospitality during research visits and acknowledges financial support from Quantum Optics Lab (Turun Yliopisto) and Dipartimento di Fisica (Universit\`a di Palermo).

\appendix*
\section{Tables}

\begin{table}[!htb]
\begin{ruledtabular}
\begin{tabular}{ c | c c c c}
\multicolumn{5}{c} {\bfseries Parties and number of seats} \\ \toprule
\multicolumn{1}{c}{} & \itshape I & \itshape II & \itshape III & \itshape IV\\ \hline
\itshape KD & 10 & 7 & 7 & 6 \\
\itshape Christian Democrats & (4.2\%)&(5.3\%) &(4.9\%) & (4.0\%) \\ \hline
\itshape KESK & 48 & 55 & 51 & 35 \\
\itshape Center Party& (22.4\%)&(24.7\%) &(23.1\%) & (15.8\%)\\ \hline
\itshape KOK & 46& 40& 50& 44\\
\itshape National Coalition Party&(21.0\%) & (18.6\%)&(22.3\%) & (20.4\%)\\ \hline
\itshape PS & 1 & 3 & 5 & 39\\
\itshape True Finns &(1.0\%) &(1.6\%) &(4.1\%) & (19.1\%)\\ \hline
\itshape RKP & 12 & 9 & 10 & 10\\
\itshape Swedish People's Party &(5.1\%) & (4.6\%)&(4.6\%) &(4.3\%) \\ \hline
\itshape SDP & 51& 53& 45& 42\\
\itshape Social Democratic Party & (22.9\%)& (24.5\%)&(21.4\%) & (19.1\%)\\ \hline
\itshape VAS & 20& 19& 17& 14\\
\itshape Left Alliance &(10.9\%) & (9.9\%)& (8.8\%)&(8.1\%) \\ \hline
\itshape VIHR & 11& 14& 15& 10\\
\itshape Green League &(7.3\%) & (8.0\%)& (8.5\%)&(7.3\%) \\ 
\end{tabular}
\end{ruledtabular}
\caption {\label{tab:parties} Political party groups and seats for each parliament term. The numbers in parentheses give percentage from total number of votes.}
\end{table}

\begin{table}[!htb]
\begin{ruledtabular}
\begin{tabular}{ l l l c c }
\multicolumn{5}{c}{\bfseries Electoral districts and number of seats}  \\ \toprule
\AA land & alan & 1 \\ 
Etel\"a-savo &  esav & 6\\ 
H\"ame &  hame & 14 \\ 
Helsinki &  hels & 21 \\  
Central Finland &  keski & 10\\ 
Kymi & kymi & 12\\
Lapland & lapp & 7\\ 
Oulu & oulu & 18\\ 
Pirkanmaa & pirk & 18\\
North Karelia &  pkar & 6\\ 
Pohjois-Savo & psav & 10\\ 
Satakunta & sata & 9\\ 
Uusimaa&  uusi &  34 \\ 
Vaasa &  vaas & 17\\ 
Varsinais-Suomi & vars &  17\\ 
\end{tabular}
\end{ruledtabular}
\caption {\label{tab:districts} Electoral districts, their abbreviations, and number of seats.}
\end{table}

\begin{table}[!htb]
\begin{ruledtabular}
\begin{tabular}{ c | c  c  c c}
\multicolumn{5}{c}{\bfseries Party coalition and position}  \\ \toprule
\multicolumn{1}{c}{}  & \itshape I & \itshape II & \itshape III & \itshape IV\\ \hline
\itshape Gov. & sdp,vas,vihr, &  kesk,sdp,rkp & kesk,rkp,kok, & sdp,vas,vihr,\\
& kok,rkp & & vihr &  kd,kok,rkp \\  \hline
\itshape Opp. & kesk,kd,ps & kok,kd,ps, & sdp,vas,kd, & kesk,ps \\
& & vas, vihr & ps & \\ \hline
\itshape Right  &  \multicolumn{4}{c}{kok, kd, ps} \\
\itshape Centre & \multicolumn{4}{c}{kesk, rkp}\\
\itshape Left & \multicolumn{4}{c}{sdp, vas, vihr}\\
\end{tabular}
\end{ruledtabular}
\caption {\label{tab:government} Government/Opposition coalitions for each term and party political position.}
\end{table}

\begin{table}[!htb]
\begin{ruledtabular}
\begin{tabular}{ c | l l l }
\multicolumn{4}{c}{\bfseries Parliament members overlap}  \\ \toprule
\multicolumn{1}{c}{}  & \itshape II & \itshape III & \itshape IV\\ \hline
\itshape I &  0.575 & 0.374 & 0.173 \\
\itshape II & - & 0.552 & 0.274 \\ 
\itshape III  &  - & - & 0.388 \\
\end{tabular}
\end{ruledtabular}
\caption {\label{tab:members} Matrix showing the overlap of members between each pair of terms, as a fraction of the maximum possible overlap, that is, the minimum number of members during both terms considered. This number is not 200 because whoever signed less than 2 initiatives was not considered.}
\end{table}

\begin{table}[!htb]
\centering
\begin{tabular}{lll}
\multicolumn{3}{c}{\bfseries Reciprocity top 3 scores}  \\ \toprule
\textit{Surname} & \textit{Party} & \textit{Signatures}  \\ \hline
\textit{Tiilikainen} & \textbf{kesk} & 3,002   \\
\textit{Ruohonen} & \textbf{ps} & 1,412    \\
\textit{Kalmari} & \textbf{kesk} & 292   \\
\end{tabular}
\caption{Reciprocity top 3 scorers. All are from opposition parties and receive the highest number of unreciprocated signatures.}
\label{tab:reciprocity}
\end{table}

\begin{table}[!htb]
\begin{ruledtabular}
\begin{tabular}{ c | c  c  c c c c}
\multicolumn{7}{c}{\bfseries Communities for the I parliament}  \\ \toprule
\multicolumn{1}{c}{} $N_C$ & \itshape Party & \itshape District & \itshape Gender & \itshape Area & \itshape Coalition & \itshape Position \\ \hline
53 & sdp, vas & pirk &- & rural & gov & left\\
48 & kesk &- &- &- & opp & centre\\
23  &  kok & uusi &- & metro & gov & right \\
15 & - & vars &- &- &- &- \\
12 & rkp & vaas &- &- &- &-\\
11 & vihr &- & f &- &- & left\\
10 & kd &- &- &- & opp & right\\
\end{tabular}
\end{ruledtabular}
\caption {\label{tab:term_I} Community characterization for the I term. $N_C$ is the number of members in each community, characterizing attributes are indicated on the top row. Communities are ordered by decreasing size.}
\end{table}

\begin{table}[!htb]
\begin{ruledtabular}
\begin{tabular}{ c | c  c  c c c c}
\multicolumn{7}{c}{\bfseries Communities for the II parliament}  \\ \toprule
\multicolumn{1}{c}{} $N_C$ & \itshape Party & \itshape District & \itshape Gender & \itshape Area & \itshape Coalition & \itshape Position \\ \hline
40 & kok & - &- & - & opp & right\\
35 & vihr & uusi &- & metro & - & left\\
26 &  vas & psav &- & - & - & left \\
25 & kesk & pkar, kymi &- & rural & gov & centre \\
18 & - & hame, vars &- &- &- &-\\
15 & ps & vaas &- &- &- & -\\
9 & - & pirk &- &- & - & -\\
5 & - & esav &- &- & - & -\\
4 & - & sata &- &- & - & -\\
\end{tabular}
\end{ruledtabular}
\caption {\label{tab:term_II} Community characterization for the II term. $N_C$ is the number of members in each community, characterizing attributes are indicated on the top row. Communities are ordered by decreasing size.}
\end{table}

\begin{table}[!htb]
\begin{ruledtabular}
\begin{tabular}{ c | c  c  c c c c}
\multicolumn{7}{c}{\bfseries Communities for the III parliament}  \\ \toprule
\multicolumn{1}{c}{} $N_C$ & \itshape Party & \itshape District & \itshape Gender & \itshape Area & \itshape Coalition & \itshape Position \\ \hline
37 & sdp & - &- & rural & opp & left\\
30 & ps & uusi &- & metro & - & -\\
22 & rkp & vaas &- & - & - & centre \\
20 & vas & - &- & - & opp & left \\
14 & - & pirk &- &- &- &-\\
13 & - & vars &- &- &- & -\\
12 & - & hame &- &- & - & -\\
10 & - & esav &- &- & - & -\\
3 & - & lapp &- &- & - & -\\
\end{tabular}
\end{ruledtabular}
\caption {\label{tab:term_III} Community characterization for the III term.  $N_C$ is the number of members in each community, characterizing attributes are indicated on the top row. Communities are ordered by decreasing size.}
\end{table}

\begin{table}[!htb]
\begin{ruledtabular}
\begin{tabular}{ c | c  c  c c c c}
\multicolumn{7}{c}{\bfseries Communities for the IV parliament}  \\ \toprule
\multicolumn{1}{c}{} $N_C$ & \itshape Party & \itshape District & \itshape Gender & \itshape Area & \itshape Coalition & \itshape Position \\ \hline
60 & sdp & - &- & - & gov & left\\
40 & ps & - &- & - & opp & right\\
40 & kesk & - &- & - & opp & centre \\
13 & rkp & vaas &- & - & - & - \\
\end{tabular}
\end{ruledtabular}
\caption {\label{tab:term_IV} Community characterization for the IV term. $N_C$ is the number of members in each community, characterizing attributes are indicated on the top row. Communities are ordered by decreasing size.}
\end{table}

\begin{table}[!htb]
\centering
\begin{tabular}{l l l l l l l r}
\multicolumn{8}{c}{\bfseries Disparity top 10 Scores}  \\ \toprule
\textbf{Surname} & \textbf{Party} & $\mathbf{n_p}$ & $\mathbf{n_s}$ & $\mathbf{n_r}$ & $\mathbf{m_s}$ & $\mathbf{m_r}$ & $\mathbf{d_1}$ \\ \hline
\textit{Tiilikainen} & \textbf{kesk} & 147 & 40 & 4750 & 26 & 39 & 0.57 \\
Thors & rkp & 22 & 15 & 31 & 10 & 4 & 0.19 \\
\textit{Ruohonen} & \textbf{ps} & 195 & 168 & 6174 & 45 & 51 & 0.07 \\
Arhinmaki & vas & 3 & 6 & 30 & 5 & 11 & -0.33 \\
Nylander & rkp & 7 & 17 & 7 & 11 & 3 & -0.42 \\
Nylund & rkp & 23 & 59 & 83 & 21 & 40 & -0.44 \\
Gestrin & rkp & 6 & 17 & 20 & 13 & 7 & -0.48 \\
Hanninen & vas & 3 & 13 & 12 & 9 & 8 & -0.62 \\
Kanerva & kok & 9 & 41 & 102 & 34 & 26 & -0.64 \\
Eloranta & sdp & 9 & 41 & 136 & 26 & 75 & -0.64 \\
 &  &  &  &  &  &  &  \\ \hline
\textbf{Surname} & \textbf{Party} & $\mathbf{n_p}$ & $\mathbf{n_s}$ & $\mathbf{n_r}$ & $\mathbf{m_s}$ & $\mathbf{m_r}$ & $\mathbf{d_2}$ \\ \hline
Harkimo & kok & 1 & 41 & 103 & 31 & 103 & 0.43 \\
Kopra & kok & 1 & 44 & 101 & 33 & 101 & 0.39 \\
Sarkomaa & kok & 1 & 27 & 56 & 19 & 56 & 0.35 \\
Makela & kok & 4 & 36 & 256 & 29 & 133 & 0.28 \\
Kymalainen & sdp & 2 & 55 & 188 & 39 & 135 & 0.26 \\
Mantymaa & kok & 1 & 72 & 120 & 33 & 120 & 0.25 \\
Arhinmaki & vas & 3 & 6 & 30 & 5 & 11 & 0.25 \\
Sinnemaki & vihr & 2 & 13 & 39 & 10 & 34 & 0.20 \\
Kataja & kok & 3 & 35 & 143 & 26 & 132 & 0.15 \\
Lapintie & vas & 1 & 27 & 31 & 13 & 31 & 0.07 \\
 &  &  &  &  &  &  &  \\ \hline
\textbf{Surname} & \textbf{Party} & $\mathbf{n_p}$ & $\mathbf{n_s}$ & $\mathbf{n_r}$ & $\mathbf{m_s}$ & $\mathbf{m_r}$ & $\mathbf{d_3}$ \\ \hline
\textit{Tiilikainen} & \textbf{kesk} & 147 & 40 & 4750 & 26 & 39 & 0.98 \\
\textit{Ruohonen} & \textbf{ps} & 195 & 168 & 6174 & 45 & 51 & 0.95 \\
Makela & kok & 4 & 36 & 256 & 29 & 133 & 0.75 \\
Arhinmaki & vas & 3 & 6 & 30 & 5 & 11 & 0.67 \\
Satonen & kok & 5 & 50 & 247 & 31 & 129 & 0.66 \\
Pelkonen & kok & 5 & 45 & 190 & 35 & 130 & 0.62 \\
Kataja & kok & 3 & 35 & 143 & 26 & 132 & 0.61 \\
Autto & kok & 4 & 57 & 225 & 39 & 136 & 0.60 \\
Kauma & kok & 4 & 37 & 131 & 28 & 89 & 0.56 \\
Tolvanen & kok & 4 & 52 & 183 & 32 & 109 & 0.56 \\
 &  &  &  &  &  &  &  \\ \hline
\textbf{Surname} & \textbf{Party} & $\mathbf{n_p}$ & $\mathbf{n_s}$ & $\mathbf{n_r}$ & $\mathbf{m_s}$ & $\mathbf{m_r}$ & $\mathbf{d_4}$ \\ \hline
Kataja & kok & 3 & 35 & 143 & 26 & 132 & 0.67 \\
Toivakka & kok & 4 & 46 & 145 & 26 & 133 & 0.67 \\
Makela & kok & 4 & 36 & 256 & 29 & 133 & 0.64 \\
Satonen & kok & 5 & 50 & 247 & 31 & 129 & 0.61 \\
\textit{Kalmari} & \textbf{kesk} & 11 & 229 & 427 & 40 & 162 & 0.60 \\
Kalliorinne & vas & 6 & 46 & 151 & 30 & 117 & 0.59 \\
Pelkonen & kok & 5 & 45 & 190 & 35 & 130 & 0.58 \\
Mantymaa & kok & 1 & 72 & 120 & 33 & 120 & 0.57 \\
Viitamies & sdp & 6 & 65 & 174 & 33 & 122 & 0.57 \\
Autto & kok & 4 & 57 & 225 & 39 & 136 & 0.55
\end{tabular}
\caption{Disparity top scorers for the IV term. Opposition parties are in bold letters. The top 3 scorers in Reciprocity, all from opposition parties, score high in disparity measures and are written in italics.}
\label{tab:disparity_top}
\end{table}

\begin{table}[!htb]
\centering
\begin{tabular}{lllllllr}
\multicolumn{8}{c}{\bfseries Disparity worst 10 Scores}  \\ \toprule
\textbf{Surname} & \textbf{Party} & $\mathbf{n_p}$ & $\mathbf{n_s}$ & $\mathbf{n_r}$ & $\mathbf{m_s}$ & $\mathbf{m_r}$ & $\mathbf{d_1}$ \\ \hline
Puumala & \textbf{kesk} & 1 & 215 & 101 & 36 & 101 & -0.99 \\
Koskela & \textbf{ps} & 3 & 528 & 53 & 56 & 28 & -0.99 \\
Vaatainen & \textbf{ps} & 1 & 368 & 25 & 54 & 25 & -0.99 \\
Vahamaki & \textbf{ps} & 3 & 455 & 30 & 58 & 10 & -0.99 \\
Alatalo & \textbf{kesk} & 2 & 208 & 14 & 47 & 11 & -0.98 \\
Lintila & \textbf{kesk} & 2 & 229 & 35 & 32 & 34 & -0.98 \\
Lohi & \textbf{kesk} & 2 & 239 & 2 & 43 & 1 & -0.98 \\
Maijala & \textbf{kesk} & 2 & 205 & 110 & 48 & 104 & -0.98 \\
Eerola & \textbf{ps} & 4 & 460 & 143 & 66 & 76 & -0.98 \\
Virtanen & \textbf{ps} & 2 & 244 & 21 & 37 & 13 & -0.98 \\
 &  &  &  &  &  &  &  \\ \hline
\textbf{Surname} & \textbf{Party} & $\mathbf{n_p}$ & $\mathbf{n_s}$ & $\mathbf{n_r}$ & $\mathbf{m_s}$ & $\mathbf{m_r}$ & $\mathbf{d_2}$ \\ \hline
Lohi & \textbf{kesk} & 2 & 239 & 2 & 43 & 1 & -0.99 \\
Korhonen & \textbf{kesk} & 13 & 226 & 33 & 47 & 14 & -0.98 \\
Pirttilahti & \textbf{kesk} & 27 & 265 & 79 & 50 & 23 & -0.98 \\
Torniainen & \textbf{kesk} & 18 & 245 & 34 & 43 & 11 & -0.98 \\
Hautala & \textbf{kesk} & 49 & 278 & 207 & 48 & 13 & -0.97 \\
Kaikkonen & \textbf{kesk} & 24 & 215 & 79 & 46 & 41 & -0.97 \\
Rundgren & \textbf{kesk} & 3 & 239 & 10 & 44 & 8 & -0.97 \\
Vehkapera & \textbf{kesk} & 6 & 223 & 18 & 43 & 10 & -0.97 \\
Niikko & \textbf{ps} & 31 & 565 & 296 & 63 & 38 & -0.97 \\
Yrttiaho & vihr & 3 & 56 & 3 & 39 & 1 & -0.96 \\
 &  &  &  &  &  &  &  \\ \hline
\textbf{Surname} & \textbf{Party} & $\mathbf{n_p}$ & $\mathbf{n_s}$ & $\mathbf{n_r}$ & $\mathbf{m_s}$ & $\mathbf{m_r}$ & $\mathbf{d_3}$ \\ \hline
Lohi & \textbf{kesk} & 2 & 239 & 2 & 43 & 1 & -0.98 \\
Jaskari & kok & 1 & 50 & 1 & 31 & 1 & -0.96 \\
Peltokorpi & \textbf{kesk} & 1 & 71 & 2 & 5 & 2 & -0.95 \\
Rundgren & \textbf{kesk} & 3 & 239 & 10 & 44 & 8 & -0.92 \\
Yrttiaho & vihr & 3 & 56 & 3 & 39 & 1 & -0.90 \\
Vahamaki & \textbf{ps} & 3 & 455 & 30 & 58 & 10 & -0.88 \\
Alatalo & \textbf{kesk} & 2 & 208 & 14 & 47 & 11 & -0.87 \\
Vaatainen & \textbf{ps} & 1 & 368 & 25 & 54 & 25 & -0.87 \\
Tiainen & vas & 1 & 64 & 5 & 44 & 5 & -0.86 \\
Vehkapera & \textbf{kesk} & 6 & 223 & 18 & 43 & 10 & -0.85 \\
 &  &  &  &  &  &  &  \\ \hline
\textbf{Surname} & \textbf{Party} & $\mathbf{n_p}$ & $\mathbf{n_s}$ & $\mathbf{n_r}$ & $\mathbf{m_s}$ & $\mathbf{m_r}$ & $\mathbf{d_4}$ \\ \hline
Lohi & \textbf{kesk} & 2 & 239 & 2 & 43 & 1 & -0.95 \\
Yrttiaho & vihr & 3 & 56 & 3 & 39 & 1 & -0.95 \\
Jaskari & kok & 1 & 50 & 1 & 31 & 1 & -0.94 \\
Feldt & sdp & 2 & 23 & 2 & 18 & 1 & -0.89 \\
Ojala & sdp & 4 & 43 & 7 & 29 & 2 & -0.87 \\
Saarinen & sdp & 6 & 47 & 11 & 32 & 3 & -0.83 \\
Palm & kd & 7 & 74 & 28 & 36 & 4 & -0.80 \\
Tiainen & vas & 1 & 64 & 5 & 44 & 5 & -0.80 \\
Mustajarvi & vihr & 13 & 65 & 16 & 46 & 5 & -0.80 \\
Jaaskelainen & kd & 6 & 59 & 24 & 25 & 4 & -0.72
\end{tabular}
\caption{Disparity worst 10 scorers for the IV term. Opposition parties are in bold letters. Lowest scores are dominated by opposition members who mainly support their spokesmen.}
\label{tab:disparity_worst}
\end{table}

\end{document}